\documentclass{article}
\usepackage{float}
\usepackage{array}     
\usepackage{tabularx}  
\usepackage[table]{xcolor} 
\usepackage{colortbl}  
\usepackage{hyperref} 
\usepackage{makecell}

\usepackage[
  paperheight=8.5in,
  paperwidth=5.5in,
  left=10mm,
  right=10mm,
  top=20mm,
  bottom=20mm]{geometry}
\usepackage[utf8]{inputenc}

\usepackage{graphicx}
\usepackage{wrapfig}
\usepackage[bottom]{footmisc}
\usepackage{listings}
\usepackage{enumitem}

\usepackage{wrapfig}
\usepackage{ragged2e}

\usepackage{array}
\usepackage[table]{xcolor}
\usepackage{multirow}
\usepackage{booktabs}
\usepackage{hhline}
\definecolor{palegreen}{rgb}{0.6,0.98,0.6}

\usepackage{amsmath}
\usepackage{amssymb}
\usepackage{multicol}
\usepackage{lipsum}
\usepackage{hyphenat}
\PassOptionsToPackage{hyphens}{url}
\usepackage{url}

\usepackage{rotating}

\usepackage[T1]{fontenc}
\usepackage{textcomp}
\usepackage{listings}
\usepackage{setspace}

\newcommand*{\email}[1]{\small{\texttt{#1}}}

\renewcommand{\footnoterule}{%
  \kern -3pt
  \hrule width \textwidth height 0.5pt
  \kern 2pt
}

\date{}

\usepackage{titlesec}
\titleformat*{\section}{\large\bfseries}
\titleformat*{\subsection}{\normalsize\bfseries}
\titleformat*{\subsubsection}{\normalsize\bfseries}

\usepackage{biblatex}

\title{A Detailed Comparative Analysis of Blockchain Consensus Mechanisms\footnote{\protect Copyright \copyright 2025 by the Consortium for Computing Sciences in Colleges.
Permission to copy without fee all or part of this material is granted provided
that the copies are not made or distributed for direct commercial advantage,
the CCSC copyright notice and the title of the publication and its date appear,
and notice is given that copying is by permission of the Consortium for
Computing Sciences in Colleges.  To copy otherwise, or to republish, requires
a fee and/or specific permission.
}
}

 \author{
 Kaeli Andrews, Linh B. Ngo, and Md Amiruzzaman\\
 Department of Computer Science\\
 West Chester University West Chester, PA 19383\\
 \email{\{ka993962, lngo, mamiruzzaman\}}{@wcupa.edu}\\
 }

\begin{document}
\maketitle

\begin{abstract}
This paper presents a comprehensive comparative analysis of two dominant blockchain consensus mechanisms—Proof of Work (PoW) and Proof of Stake (PoS)—evaluated across seven critical metrics: energy use, security, transaction speed, scalability, centralization risk, environmental impact, and transaction fees. Utilizing recent academic research and real-world blockchain data, the study highlights that PoW offers robust, time-tested security but suffers from high energy consumption, slower throughput, and centralization through mining pools. In contrast, PoS demonstrates improved scalability and efficiency, significantly reduced environmental impact, and more stable transaction fees; however, it raises concerns over validator centralization and long-term security maturity. The findings underscore the trade-offs inherent in each mechanism and suggest hybrid designs may combine PoW’s security with PoS’s efficiency and sustainability. The study aims to inform future blockchain infrastructure development by striking a balance between decentralization, performance, and ecological responsibility.

\end{abstract}

\section{Introduction}
Public discourse often reduces blockchain to speculation and cryptocurrency schemes \cite{mckenzie2023easy}, obscuring its broader role as a decentralized infrastructure for trustless collaboration. Blockchain technology functions as a distributed digital ledger maintained by a network of computers that reach consensus on every update \cite{zheng2017overview, balaji2017decentralization}. This architecture enables transparent, trustless interaction and supports applications beyond cryptocurrency, such as supply chain management \cite{ashishkumarjha2022supplychain}, digital identity, and voting systems \cite{sarmah2018understanding}. Blockchain represents a foundational shift in digital trust by enabling value transfer, record verification, and transaction execution without intermediaries.

Blockchain is often celebrated for its transparency and security, but it is ultimately the consensus mechanism that determines whether those ideals are truly achieved. A consensus mechanism is defined as the protocol through which nodes reach agreement on the validity of transactions. The two most widely adopted consensus mechanisms are Proof of Work (PoW) and Proof of Stake (PoS), which currently power leading cryptocurrencies and support a range of decentralized applications across industries \cite{sarmah2018understanding, ashishkumarjha2022supplychain}. PoW and PoS determine who writes the next \emph{page} in the blockchain, the system’s overall security \cite{bonneau2015sok, li2020security}, and its associated energy costs \cite{emmanuel2025energy, zimba2025energy}. Comparing PoW and PoS is essential for understanding blockchain’s broader societal, economic, and environmental impacts \cite{jha2024powvpos}. This comparison demands academic rigor and practical insight to capture the nuanced implications of each mechanism.

This study conducts a meta-analysis that compares the technical foundations and real-world implications of PoW and PoS. It breaks down how each mechanism functions, identifies major cryptocurrencies and platforms that rely on them, and weighs their respective strengths and limitations. Furthermore, the analysis examines the broader implications of each mechanism by assessing who benefits, who is excluded, and how each mechanism shapes future blockchain infrastructure in terms of energy consumption, decentralization, and long-term viability \cite{jha2024powvpos, sarmah2018understanding}. The specific contributions of this study are as follows.

\begin{itemize}
    \item A detailed comparative analysis of \textbf{Proof of Work (PoW)} and \textbf{Proof of Stake (PoS)}, highlighting their technical, economic, and environmental trade-offs.
    \item Quantitative evaluation of \textbf{centralization risks} across PoW and PoS networks, including analysis of \textbf{validator} and \textbf{mining pool} concentration.
    \item Empirical evidence of \textbf{PoW's high energy consumption} versus \textbf{PoS's energy efficiency}, with real-world examples (e.g., Bitcoin vs. Ethereum post-Merge).
    \item Analysis of \textbf{scalability limitations} in PoW (slow throughput) versus PoS (higher transaction speeds), supported by network data.
    \item Evaluation of \textbf{societal implications}, including governance fairness, wealth concentration in PoS, and environmental sustainability.
    \item Insights into \textbf{future trends}, such as hybrid consensus models, to balance security, decentralization, and efficiency.
\end{itemize}

The remainder of this paper is organized as follows. Section \ref{sec:review} briefly discusses previous foundational technical work for blockchain technology and the two consensus mechanisms, PoW and PoS. Section \ref{sec:method} describes our comparison methodology. Section \ref{sec:result} presents the comparison and analyzes the implications. In Section \ref{sec:conclusion}, we conclude the paper, identify limitations, and suggest future work. 
 
\section{Literature Review}
\label{sec:review}
Blockchain networks rely on consensus mechanisms to verify transactions and maintain decentralization. Proof of Work (PoW) and Proof of Stake (PoS) are the two most studied models, each with trade-offs across energy use, security, speed, scalability, centralization, environmental impact, and transaction fees.

\textbf{Energy use} is a defining distinction between PoW and PoS. PoW relies on energy-intensive mining, with Bitcoin’s electricity consumption rivaling that of entire nations \cite{zimba2025energy, yang2023blockchain, gervais2016pow}. This demand grows with network difficulty, raising sustainability concerns \cite{emmanuel2025energy}. PoS avoids mining altogether, runs on standard infrastructure, and reduces energy use by over 99 percent, as seen in Ethereum (post-Merge) \cite{zimba2025energy}.

\textbf{Security} in PoW derives from computational difficulty and cost, deterring Sybil and 51 percent attacks through resource expenditure \cite{bonneau2015sok, zheng2017overview, gervais2016pow}. PoS relies on financial penalties and stake-based selection, with mechanisms like slashing used to enforce validator honesty \cite{yakubu2024consensus, castro2024unsealing, kiayias2017ouroboros}.

\textbf{Speed and throughput} vary significantly. PoW networks like Bitcoin process blocks slowly with limited transaction rates \cite{bonneau2015sok}. PoS systems, including Ethereum (post-Merge) and Polkadot, offer faster confirmation and higher throughput \cite{li2020security, jha2024powvpos}.

\textbf{Scalability} is limited in PoW due to strict consensus requirements, often addressed through external solutions. PoS integrates scalability more natively through designs like Hydra and parachains \cite{ankit2024survey, bulgakov2024scalability, chen2024scalability}.

\textbf{Centralization risk} emerges from different pressures. PoW networks see mining pool dominance, often reflected in low Nakamoto Coefficient scores \cite{nakaflow2025, ccn2025nakamoto}. In PoS, wealth concentration and custodial staking services threaten validator diversity, though some networks like Cardano demonstrate mitigation efforts \cite{brookings2023centralization, natoli2019survey}.

\textbf{Environmental impact} is significant in PoW due to electricity and e-waste from mining hardware \cite{rmi2023energy, yang2023blockchain}. PoS drastically lowers these costs by design \cite{zimba2025energy}.

\textbf{Transaction fees} are often volatile in PoW during congestion due to fixed block capacity \cite{bonneau2015sok, jain2024fees}. PoS networks typically implement adaptive fee models to maintain fee stability and performance under load \cite{inami2025queueing}.

This review establishes the foundation for our comparative analysis. The next section details the methodology used to evaluate PoW and PoS across these seven metrics, followed by a results-driven comparison using current examples and empirical data.

\section{Methodology}
\label{sec:method}

This study adopted an approach similar to meta-analysis to analyze the two major blockchain consensus mechanisms: Proof of Work (PoW) and Proof of Stake (PoS). We first discuss the fundamental designs of blockchain's operations and consensus mechanisms, then explain how these designs inform our choices of comparative metrics.

\subsection{Fundamental designs}
\begin{figure}[t]
\centering
\includegraphics[width=0.8\textwidth]{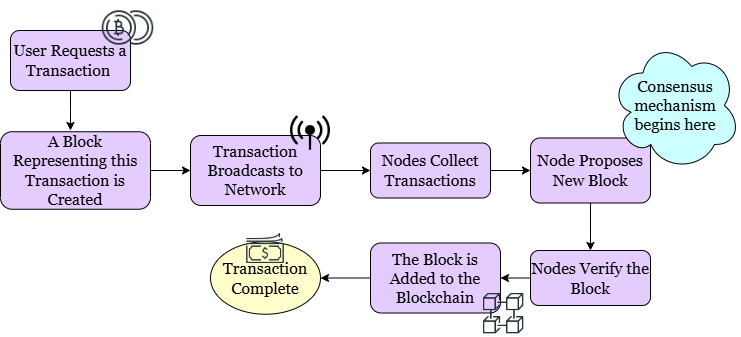}
\caption{\label{fig:blockchain}Overview of Blockchain Technology Flowchart}
\end{figure}

Blockchain is a decentralized digital ledger that records transactions securely and transparently across a distributed network of computers \cite{zheng2017overview, balaji2017decentralization}. As illustrated in Figure \ref{fig:blockchain}, a transaction begins when a user broadcasts it to the network. Next, network nodes verify the transaction's validity, group it into a block, and append it to the chain through a consensus mechanism \cite{sarmah2018understanding, ashishkumarjha2022supplychain}. Each block contains a cryptographic hash of the previous block, ensuring that the data remains immutable \cite{zheng2017overview}. The structure of blockchain eliminates the need for centralized control, making it foundational to systems that require trust, transparency, and resilience \cite{yan2022centralization, balaji2017decentralization}.

\begin{figure}[t]
\centering
\includegraphics[width=0.8\textwidth]{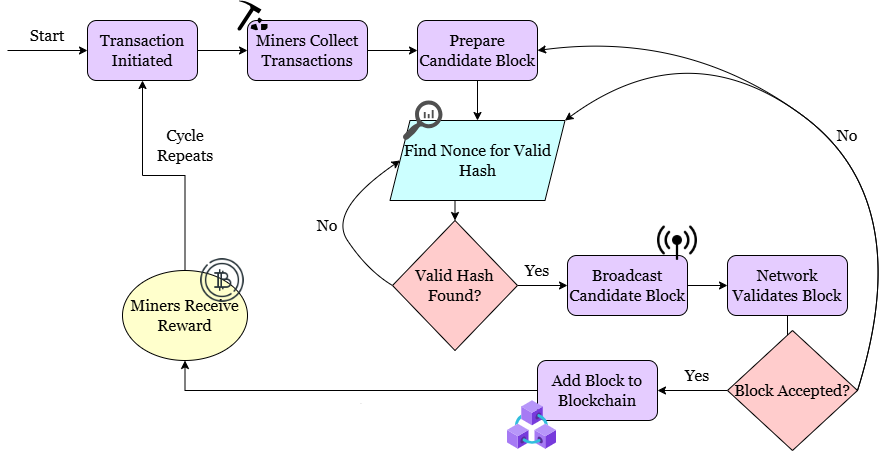}
\caption{\label{fig:pow}Proof of Work Flowchart}
\end{figure}

The PoW consensus mechanism, shown in Figure \ref{fig:pow}, secures the blockchain through computational competition \cite{bonneau2015sok}. Miners build candidate blocks from pending transactions and repeatedly hash each block's data with nonce values until the resulting hash meets the network's difficulty target \cite{yan2022centralization, gervais2016pow}. This target regulates how often new blocks are added, adjusting automatically to maintain a consistent block time, such as every 10 minutes in Bitcoin \cite{gervais2016pow, jha2024powvpos}. The process is intentionally resource-intensive, making it difficult for any single actor to manipulate the ledger \cite{bonneau2015sok}.

\begin{figure}[t]
\centering
\includegraphics[width=0.8\textwidth]{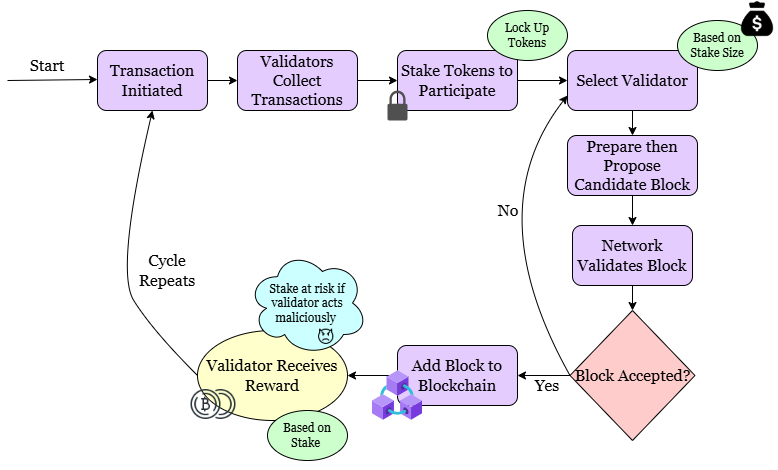}
\caption{\label{fig:pos}Proof of Stake Flowchart}
\end{figure}

The PoS consensus mechanism, shown in Figure \ref{fig:pos}, secures the blockchain through stake-based validator selection. Validators are selected to propose new blocks based on the amount of cryptocurrency they lock as collateral \cite{bentov2014cryptocurrencies}, with weighted, pseudo-random algorithms favoring larger stakes \cite{bentov2014activity}. This process is formalized in protocols such as Ouroboros \cite{kiayias2017ouroboros} and evaluated in comparative studies \cite{jha2024powvpos}. The selected validator assembles pending transactions into a block and broadcasts it to the network, where other validators vote on its validity \cite{yan2022centralization}. If approved, the block is added and rewards are distributed in transaction fees or new tokens \cite{jha2024powvpos}. To deter dishonest behavior, PoS networks enforce slashing penalties that remove part of a validator’s stake for submitting invalid or conflicting blocks \cite{kiayias2017ouroboros, bentov2014cryptocurrencies, bentov2014activity, yan2022centralization}.

\subsection{Metric Selections}

As shown in Table \ref{tab:metrics}, the analysis focused on several core evaluation metrics commonly discussed in blockchain research.

\section{Results}
\label{sec:result}

Our analysis compares Proof of Work (PoW) and Proof of Stake (PoS) using key evaluation metrics, as summarized in Table \ref{tab:metrics}. Each metric highlights a different aspect of how consensus mechanisms shape blockchain systems and their broader impact. 
The following subsections present the results for each category in turn.

\begin{table}[t]
\centering
\caption{\label{tab:metrics}Key Evaluation Metrics for Blockchain Consensus Mechanisms}
\scalebox{0.85}{ 
\footnotesize 
\begin{tabular}{l l l} 
\toprule
\rowcolor[HTML]{C0F0A3} 
\textbf{Metric} & \textbf{Description} & \textbf{Papers} \\
\midrule
\makecell[l]{Energy Use} & \makecell[l]{Examines the computational and environmental costs \\ associated with each mechanism.} & \makecell[l]{\cite{bentov2014cryptocurrencies, bentov2014activity, gervais2016pow},\\ \cite{jha2024powvpos, emmanuel2025energy, rmi2023energy, zimba2025energy,yang2023blockchain}} \\ 
\midrule
\makecell[l]{Security} & \makecell[l]{Assesses resilience against common attacks and \\ network vulnerabilities.} & \makecell[l]{\cite{yakubu2024consensus, bonneau2015sok, castro2024unsealing, gervais2016pow, jha2024powvpos},\\ \cite{kiayias2017ouroboros, li2020security, bulgakov2024scalability, yan2022centralization}} \\ 
\midrule
\makecell[l]{Speed and \\ Throughput} & \makecell[l]{Evaluates how quickly transactions are processed \\ and confirmed.} & \makecell[l]{\cite{bonneau2015sok, jha2024powvpos, li2020security, bulgakov2024scalability, zheng2017overview}} \\ 
\midrule
\makecell[l]{Scalability} & \makecell[l]{Determines each mechanism's ability to handle \\ increasing network demands.} & \makecell[l]{\cite{bonneau2015sok, bulgakov2024scalability, chen2024scalability},\\ \cite{ankit2024survey, jha2024powvpos, li2020security, zheng2017overview}} \\ 
\midrule
\makecell[l]{Centralization \\ Risk} & \makecell[l]{Analyzes the extent of power concentration within a network, \\ using the \emph{Nakamoto Coefficient} to quantify the number \\ of entities required to compromise consensus.} & \makecell[l]{\cite{balaji2017decentralization, bentov2014activity, bentov2014cryptocurrencies},\\ \cite{bonneau2015sok, brookings2023centralization, ccn2025nakamoto},\\ \cite{coinstats2025nakamoto, kiayias2017ouroboros, li2020security, nakaflow2025, yan2022centralization}} \\ 
\midrule
\makecell[l]{Environmental \\ Impact} & \makecell[l]{Reviews broader ecological effects beyond raw energy usage.} & \makecell[l]{\cite{jha2024powvpos, jain2024fees, rmi2023energy, yang2023blockchain, zimba2025energy}} \\ 
\midrule
\makecell[l]{Transaction Fees} & \makecell[l]{Considers how network congestion affects transaction costs.} & \makecell[l]{\cite{jha2024powvpos, li2020security, inami2025queueing,  yan2022centralization, jain2024fees}} \\ 
\bottomrule
\end{tabular}
} 
\end{table}

\subsection{Energy Use}
Energy use refers to the amount of electricity consumed by a blockchain network to maintain consensus and secure the system. Proof of Work (PoW) consumes significantly more energy than Proof of Stake (PoS) due to its reliance on computational mining \cite{gervais2016pow}. PoW requires miners to perform brute-force computations to solve cryptographic puzzles, which consumes large amounts of electricity \cite{jha2024powvpos}. Bitcoin mining alone consumes an estimated 100–150 terawatt-hours per year \cite{emmanuel2025energy}, exceeding the annual energy consumption of countries such as Norway \cite{yang2023blockchain, rmi2023energy}. This level of demand raises substantial environmental and economic concerns about the sustainability of PoW \cite{zimba2025energy}.

PoW's design ties network security to resource expenditure, making it inherently energy-intensive. Studies show that as mining difficulty increases, energy use scales proportionally, compounding environmental impact \cite{bentov2014activity, gervais2016pow}. The development of specialized mining hardware and large-scale mining operations further intensifies carbon emissions and regional energy strain \cite{zimba2025energy, bentov2014cryptocurrencies}.

PoS eliminates the need for computational competition by selecting validators based on their stake in the network \cite{jha2024powvpos}. This model significantly reduces energy consumption, as validators only maintain network connectivity and use digital signatures to approve new blocks \cite{emmanuel2025energy}. After Ethereum’s 2022 transition to PoS, its energy use dropped to an estimated 0.01 terawatt-hours per year \cite{yang2023blockchain}. Sustainability assessments confirm that PoS architectures consume dramatically less energy than PoW systems, often by several orders of magnitude \cite{emmanuel2025energy}.

\subsection{Security}
Security refers to a blockchain network's ability to resist attacks, maintain data integrity, and ensure that only valid transactions are confirmed. Proof of Work (PoW) and Proof of Stake (PoS) approach security through different cost structures and deterrents.

In PoW, miners compete to solve cryptographic puzzles, with difficulty adjusted dynamically to maintain an average block time of around 10 minutes. This prevents blocks from being created too quickly and keeps the network synchronized \cite{gervais2016pow}. The high computational and hardware costs required to control a majority of the network’s hash power make 51 percent attacks economically unfeasible at scale \cite{zheng2017overview, li2020security}. PoW is also resistant to Sybil attacks, as creating many fake identities requires significant computational resources \cite{bonneau2015sok}. Over time, PoW has demonstrated real-world robustness. Bitcoin, for instance, has operated securely for over a decade without a successful consensus-level attack \cite{gervais2016pow, li2020security}.

PoS secures the network by selecting validators based on their stake in the protocol \cite{jha2024powvpos}. Validators with greater holdings have more at stake, incentivizing honest behavior \cite{kiayias2017ouroboros}. Malicious actions such as signing conflicting blocks or altering transaction data can result in slashing penalties that destroy a portion of the validator's stake \cite{li2020security}. This introduces direct financial penalties, aligning network security with validator incentives.

However, PoS introduces different vulnerabilities. Because block production is inexpensive, validators might attempt to sign multiple competing chains, known as the Nothing-at-Stake problem \cite{castro2024unsealing}. Most PoS protocols counter this with slashing and finality rules, but their long-term effectiveness continues to be evaluated \cite{yakubu2024consensus}. Another concern is stake centralization. If a small number of participants accumulate most of the staked tokens, they could launch stake-based 51 percent attacks or disproportionately influence governance decisions \cite{bulgakov2024scalability}.

Although platforms like Ethereum (post-Merge) and Cardano have adopted PoS, it remains a relatively newer model. Its security relies more heavily on economic assumptions and continues to be tested under real-world conditions. In contrast, PoW is a mature and well-tested system that has demonstrated resilience in hostile environments over time \cite{kiayias2017ouroboros, castro2024unsealing}.

\subsection{Speed and Transaction Throughput}
Speed and transaction throughput refer to the rate at which a blockchain can process and confirm transactions over time. In Proof of Work (PoW) networks, speed is constrained by computational mining and fixed block intervals. Bitcoin produces blocks approximately every 10 minutes and supports an average throughput of about seven transactions per second \cite{bonneau2015sok, zheng2017overview}. Even in faster PoW systems, confirmation delays and multi-block finality reduce effective throughput \cite{li2020security, jha2024powvpos}. These structural limitations stem from the need to maintain decentralized synchronization and secure consensus.

Proof of Stake (PoS) networks remove the computational burden of mining, enabling shorter block times and higher transaction throughput. Platforms such as Ethereum (post-Merge), Cardano, and Polkadot report transaction rates ranging from 250 to over 1,000 per second, depending on their architecture and scaling strategies \cite{bulgakov2024scalability, jha2024powvpos}. PoS networks also achieve faster finality, with many confirming transactions within seconds. These characteristics make PoS well-suited for high-volume applications like decentralized finance and real-time smart contract execution \cite{jha2024powvpos, li2020security}.

While PoW's block intervals limit throughput, PoS achieves higher transaction rates through quicker finality and more efficient validation processes \cite{jha2024powvpos, bulgakov2024scalability}.

\subsection{Scalability}
Scalability refers to a blockchain's ability to maintain performance as demand increases. In Proof of Work (PoW), scalability is limited by protocol constraints and the requirement for global consensus. These factors reduce throughput and necessitate external solutions to handle increased demand \cite{bonneau2015sok, zheng2017overview}. Bitcoin and Dogecoin illustrate this challenge, processing approximately seven and thirty-three transactions per second, respectively. These rates are insufficient for large-scale applications without support from off-chain tools such as Layer 2 networks \cite{bulgakov2024scalability, ankit2024survey}.

Proof of Stake (PoS) enhances scalability by removing the need for energy-intensive mining. Validators are selected based on stake rather than computation, which lowers overhead and enables more efficient coordination \cite{jha2024powvpos, bulgakov2024scalability}. This allows PoS networks to process higher transaction volumes without overwhelming the system. Ethereum (post-Merge), for instance, aims to support up to 100,000 transactions per second using scaling mechanisms such as sharding and rollups \cite{ankit2024survey, chen2024scalability}. Sharding divides the blockchain into smaller segments, or “shards,” that process transactions in parallel to increase overall system throughput \cite{chen2024scalability, bulgakov2024scalability}. Cardano employs Hydra, an off-chain transaction protocol that reduces demand on the main layer \cite{jha2024powvpos}. Polkadot uses parachains, which are independent blockchains running in parallel with the core network, allowing the system to handle more transactions simultaneously \cite{ankit2024survey, bulgakov2024scalability}.

PoW’s architecture imposes inherent scalability limitations, requiring external solutions to maintain performance. PoS architectures often integrate native features that support higher throughput \cite{zheng2017overview, li2020security}.

\subsection{Centralization Risk}
Centralization risk refers to the concentration of control within a small number of entities capable of influencing or compromising consensus. In Proof of Work (PoW), competitive mining often leads to the formation of large mining pools. These pools aggregate computational power to improve reward probability but concentrate control among a few dominant actors \cite{yan2022centralization}. This increases the likelihood of coordinated attacks, including 51 percent attacks, where entities controlling most of the hash power can alter transactions or block new ones from being added \cite{bonneau2015sok}.

The \emph{Nakamoto Coefficient} is a common measure of decentralization. It estimates how many distinct entities must collude to disrupt consensus \cite{balaji2017decentralization}. In Bitcoin, this number is typically around 2 to 3, meaning just a few mining pools could compromise the network \cite{nakaflow2025}. Similar patterns are seen in Litecoin and Dogecoin, which share infrastructure and exhibit comparable concentration \cite{brookings2023centralization}.

PoW networks face centralization through mining pools. Proof of Stake (PoS) networks, in contrast, encounter distinct risks tied to stake distribution. Validators with larger stakes are more likely to be selected to produce blocks, giving disproportionate influence to high-stakes participants \cite{li2020security}. Many users delegate tokens to third-party services such as Lido or centralized exchanges, which operate validator nodes on their behalf \cite{bentov2014cryptocurrencies}. These services often control a significant share of the total stake and collect a commission from staking rewards \cite{brookings2023centralization}. Although delegation increases accessibility, it can centralize voting power among a small number of platforms \cite{yakubu2024consensus}. Some protocols attempt to mitigate this. Cardano's Ouroboros, for example, uses randomized leader selection and stake distribution strategies to promote broader participation and reduce wealth concentration \cite{kiayias2017ouroboros}.

In Ethereum (post-Merge), the Nakamoto Coefficient is also estimated at 2 to 3, reflecting stake concentration among validators \cite{ccn2025nakamoto}. In contrast, Cardano maintains a coefficient of about 25, indicating a more distributed set of validators \cite{coinstats2025nakamoto}.

Overall, PoW is susceptible to mining pool consolidation, while PoS can suffer from validator centralization driven by stake distribution. The Nakamoto Coefficient offers a practical way to evaluate decentralization across both systems \cite{balaji2017decentralization, brookings2023centralization}.

\subsection{Environmental Impact}
Environmental impact refers to the ecological effects of running a blockchain network, including energy consumption, hardware waste, and emissions. While both Proof of Work (PoW) and Proof of Stake (PoS) secure decentralized networks, their environmental footprints differ significantly in scale and nature.

PoW’s high energy use results from its reliance on computational mining. Miners compete to solve cryptographic puzzles, consuming a significant amount of electricity in the process. This is an intentional design feature that ties security to resource expenditure \cite{bonneau2015sok, kiayias2017ouroboros}. Bitcoin alone consumes an estimated 100–150 terawatt-hours annually, more than countries such as Argentina or Norway \cite{rmi2023energy, jain2024fees}. Specialized mining hardware like ASICs further intensifies the impact, producing electronic waste and emissions due to short life cycles and high power demands \cite{zimba2025energy}. Environmental impacts vary based on regional energy profiles. In areas with fossil-fuel-heavy energy grids, carbon intensity is higher. In some cases, coal plants have been reactivated to meet mining demand \cite{yang2023blockchain, zimba2025energy}. Subsidized electricity in certain regions also contributes to PoW’s environmental burden.

PoS significantly reduces energy use by replacing mining with stake-based validation. Validators are selected based on the amount of cryptocurrency they stake, eliminating the need for repetitive computation \cite{jha2024powvpos}. PoS networks run on standard server infrastructure. Following Ethereum's 2022 transition to PoS, energy consumption dropped by over 99 percent \cite{emmanuel2025energy}. PoS also reduces hardware turnover and electronic waste, which aligns better with renewable energy. Validator nodes can run efficiently in data centers or decentralized setups powered by clean energy \cite{zimba2025energy}. Although large custodial staking platforms may increase overall energy use, the impact remains minimal compared to PoW and can be mitigated through sustainable hosting practices.

\subsection{Transaction Fees}
Transaction fees refer to the costs users pay to have their transactions processed, often rising during periods of network congestion. Fee structures vary by consensus model, particularly between Proof of Work (PoW) and Proof of Stake (PoS).

In PoW networks like Bitcoin, transaction fees incentivize miners alongside block rewards \cite{bonneau2015sok, li2020security}. When activity increases and block space is limited, users compete by offering higher fees, creating an auction-like model. This leads to fee volatility, especially during market surges, where smaller transactions may become impractical \cite{jha2024powvpos, li2020security}. These fluctuations reflect broader scalability limitations in PoW systems, where block size and timing constraints reduce throughput and increase congestion \cite{yan2022centralization}.

PoS networks typically offer more stable and predictable fee structures. Because validators do not incur high energy costs, they rely less on transaction fees as a primary incentive \cite{jha2024powvpos}. Some networks, such as Ethereum (post-Merge), employ base fee models that adjust dynamically in response to network demand, thereby helping to reduce volatility while managing congestion \cite{jain2024fees}. Analytical models based on queueing theory suggest PoS systems experience less fee inflation under load due to faster finality and flexible confirmation processes \cite{inami2025queueing}. These efficiencies improve performance during peak usage and support high-throughput applications such as decentralized finance.

Overall, PoW networks tend to have higher and more volatile fees due to mining incentives and block constraints, while PoS networks use adaptive models that help stabilize costs during congestion.

\subsection{Final Results}

\begin{table}[t]
\caption{\label{tab:compare}Comparison of PoW and PoS in terms of features}
\begin{center}
\begin{tabular}{l l l} 
\toprule 
\rowcolor[HTML]{C0F0A3} 
\textbf{Feature} & \textbf{PoW} & \textbf{PoS} \\ 
\midrule 
Energy Use & High & Low \\ 
Security & Strong & Strong \\ 
Speed & Slow & Fast \\ 
Scalability & Limited & Scalable \\ 
Centralization Risk & High & Moderate \\ 
Environmental Impact & High & Low \\ 
Transaction Fees & Variable & Stable \\ 
\bottomrule 
\end{tabular}
\end{center}
\end{table}

Table \ref{tab:compare} summarizes the comparative outcomes of PoW and PoS. These ratings reflect the analysis across Sections 4.1–4.7. 

PoW earns a \emph{high} energy use rating due to its reliance on energy-intensive mining, while PoS earns a \emph{low} rating for removing the need for computational competition. Both mechanisms earned \emph{strong} security marks: PoW for its decade-long operational resilience, PoS for enforcing economic penalties despite shorter track records.

PoW was rated \emph{slow} and \emph{limited} in scalability due to fixed block intervals and protocol rigidity. PoS earned \emph{fast} and \emph{scalable} ratings based on faster finality and support for various optimization methods. PoW’s \emph{high} centralization risk stems from mining pool dominance, whereas PoS was rated \emph{moderate} due to validator stake concentration, with Cardano as a more decentralized example.

PoW’s \emph{high} environmental impact rating results from its energy demands and hardware waste. PoS scored \emph{low} for operating efficiently on standard infrastructure. PoW’s fees were rated as \emph{variable} due to congestion spikes, while PoS achieved \emph{stable} ratings for adaptive fee models and lower operational costs.

\section{Conclusion}
\label{sec:conclusion}
This study compared Proof of Work (PoW) and Proof of Stake (PoS) across seven metrics: energy use, security, speed, scalability, centralization risk, environmental impact, and transaction fees. PoW delivers proven security, but incurs high energy costs, slower throughput, and centralized mining dynamics. PoS improves efficiency and scalability while reducing energy use, though it raises concerns about validator concentration and long-term resilience.

Neither mechanism fully meets all technical and structural requirements. PoW is well-suited for systems that prioritize security and immutability, while PoS aligns with applications emphasizing performance and sustainability. Both involve trade-offs between decentralization, efficiency, and resilience.

Hybrid consensus models present a potential path forward by integrating the robustness of PoW with the adaptability of PoS \cite{yakubu2024consensus, ankit2024survey, bentov2014activity}. Nervos exemplifies a layered hybrid approach by using PoW for base security while enabling energy-efficient, PoS-driven smart contracts and scalability \cite{nervos2023architecture}. Future research should explore validator equity, ecological sustainability, and governance frameworks that maintain decentralized integrity at scale \cite{zimba2025energy, emmanuel2025energy}. While this analysis aligns with current literature and reported data, it does not include empirical testing and may generalize trends that differ across implementations. Despite these limitations, the comparative framework remains a valuable tool for guiding sustainable and equitable blockchain innovation.

\printbibliography

\end{document}